\documentclass[dvipdfm,twocolumn,showpacs,aps,prb]{revtex4}

\usepackage{amssymb,amsmath}
\usepackage[dvips]{graphicx}

\usepackage[dvipdfm]{hyperref}
\hypersetup{breaklinks=true, colorlinks=true, citecolor=blue, linkcolor=blue,
            filecolor=blue, urlcolor=blue}

\begin{document}

\title{Thermal expansion and Gr\"{u}neisen parameter
       in quantum Griffiths phases}

\author{Thomas Vojta}
\affiliation{Department of Physics, Missouri University of Science and Technology, Rolla, MO 65409, USA}
\affiliation{Max-Planck-Institute for Physics of Complex Systems, Noethnitzer Str. 38, 01187
Dresden, Germany}

\begin{abstract}
We consider the behavior of the Gr\"{u}neisen parameter, the ratio between thermal
expansion and specific heat, at pressure-tuned infinite-randomness quantum-critical
points and in the associated quantum Griffiths phases. We find that the Gr\"{u}neisen
parameter diverges as $\ln(T_0/T)$ with vanishing temperature $T$ in the quantum
Griffiths phases. At the infinite-randomness critical point itself, the Gr\"{u}neisen
parameter behaves as $[\ln(T_0/T)]^{1+1/(\nu\psi)}$ where $\nu$ and $\psi$ are the
correlation length and tunneling exponents. Analogous results hold for the magnetocaloric
effect at magnetic-field tuned transitions. We contrast clean and dirty systems, we
discuss subtle differences between Ising and Heisenberg symmetries, and we relate our
findings to recent experiments on CePd$_{1-x}$Rh$_x$.
\end{abstract}

\date{\today}

\maketitle


\emph{Introduction}---Quantum phase transitions (QPTs) occur at zero temperature when a
parameter such as pressure or magnetic field is varied. At these transitions, the quantum
fluctuations associated with the competition between different quantum ground states lead
to unconventional thermodynamic and transport
properties.\cite{SGCS97,Sachdev_book99,Vojta_review00,VojtaM03} In metallic systems they
can induce, e.g., non-Fermi liquid behavior and exotic superconductivity. The
characterization of QPTs is a topic of great current interest with many fundamental
questions remaining unresolved.

Over the last few years, the Gr\"{u}neisen parameter $\Gamma$, the ratio between thermal
expansion coefficient and specific heat, has become a valuable tool for analyzing
pressure-tuned QPTs. For transitions tuned by magnetic field, the same role is played by
the magnetocaloric effect. Zhu et al.\cite{ZGRS03,GarstRosch05} showed that the thermal
expansion coefficient is more singular than the specific heat at a generic clean quantum
critical point (QCP). They thus predicted that the Gr\"{u}neisen parameter diverges when
approaching criticality. Specifically, if hyperscaling holds (below the upper critical
dimension), $\Gamma \sim T^{-1/(z\nu)}$ if the temperature $T$ is lowered at the critical
pressure $p_c$ and $\Gamma \sim 1/(p-p_c)$ if the pressure $p$ approaches $p_c$ at zero
temperature ($z$ denotes the dynamical exponent).
Above the upper critical dimension, $\Gamma$ still diverges, but the
functional form is modified by dangerously irrelevant variables. Diverging Gr\"{u}neisen
parameters have since been observed at
several\cite{Kuechleretal03,Kuechleretal04,Kuechleretal06,TRGSG09} magnetic QCPs.

Since many materials feature considerable amounts of quenched randomness,
the study of QPTs in random systems has received
much attention recently. The interplay
between quantum fluctuations and static random fluctuations results in more dramatic
disorder effects at QPTs than at classical
transitions, including quantum Griffiths
singularities,\cite{ThillHuse95,GuoBhattHuse96,RiegerYoung96,CastroNetoJones00,VojtaSchmalian05}
activated dynamical scaling,\cite{Fisher92,Fisher95,HoyosKotabageVojta07}
and smeared transitions.\cite{Vojta03a,HoyosVojta08}
A review of some of these phenomena can be found in Ref.\ \onlinecite{Vojta06}.
In view of the insight about the character of a QPT
that can be gained from the Gr\"{u}neisen parameter and the
magnetocaloric effect, it is desirable to determine their behavior within
these unconventional scenarios. This is particularly timely because exotic scaling
behavior compatible with many predictions of the quantum Griffiths scenario
has  recently been observed\cite{SWKCGG07,Westerkampetal09} at the ferromagnetic QPT
in CePd$_{1-x}$Rh$_x$.

In this paper, we therefore calculate the thermal expansion coefficient and the
Gr\"{u}neisen parameter (for pressure-tuned transitions)
as well as the magnetocaloric effect (for  magnetic-field tuned transitions)
at infinite-randomness QCPs and in the associated quantum
Griffiths phases. We use two methods, a heuristic rare region theory and
a scaling analysis of the QCP itself.


We define the Gr\"{u}neisen parameter\cite{Grueneisen12,ZGRS03}
$\Gamma$ as the ratio between the thermal volume expansion
coefficient
\begin{equation}
\beta=  V^{-1} \left( {\partial V}/{\partial T} \right)_p =
      -V^{-1} \left(  {\partial S}/{\partial p} \right)_T~,
\label{eq:beta_def}
\end{equation}
and the molar specific heat
\begin{equation}
c_p = T N^{-1} \left( {\partial S}/{\partial T} \right)_p~.
\label{eq:c_def}
\end{equation}
Here, $V$ is the volume, $N$ is the particle number, and $S$ denotes the entropy.
Thus,
\begin{equation}
\Gamma = \frac \beta {c_p} = -\frac {(\partial S/\partial p)_T}{V_m T(\partial S/\partial T)_p}
\label{eq:Gamma_def}
\end{equation}
with $V_m=V/N$ the molar volume. For a pressure-tuned transition, $(\partial S /\partial
p)_T = p_c^{-1} (\partial S /\partial r)_T$ explores the dependence of the entropy on the
dimensionless distance from criticality, $r=(p-p_c)/p_c$. For a transition tuned by
magnetic field $H$ with $r=(H-H_c)/H_c$, the same dependence is encoded in
$(\partial S /\partial H)_T= (\partial M / \partial T)_H$ with $M$ the total magnetization.
We thus define the magnetic analog of the Gr\"{u}neisen parameter,
\begin{eqnarray}
\Gamma_H = -\frac {(\partial M/\partial T)_H} {c_H} = -\frac {(\partial S/\partial H)_T}{T(\partial S/\partial T)_H}
         = \frac 1 T \left(\frac {\partial T}{\partial H}\right)_S
\label{eq:Gamma_H_def}
\end{eqnarray}
which can be determined from the magnetocaloric effect.


\emph{Rare region theory}---For definiteness,
we consider a $d$-dimensional quantum Landau-Ginzburg-Wilson (LGW) free energy functional for
an $n$-component order parameter field $\phi$. The action of the clean system
is given by
\cite{Hertz76}
\begin{equation}
S=\int dx\,dy\ \phi (x)\,K (x,y)\,\phi (y)+ u \int dx\ \phi^{4}(x)~.
\label{eq:action}
\end{equation}%
Here, $x\equiv (\mathbf{x},\tau )$ comprises position $\mathbf{x}$ and
imaginary time $\tau $, and $\int dx\equiv \int d\mathbf{x}%
\int_{0}^{1/T}d\tau $.  The Fourier transform of the bare inverse propagator
(two-point vertex) $K (x,y)$ reads
$K (\mathbf{q},\omega _{n})=(r_{0}+\mathbf{q}^{2}+\gamma |\omega
_{n}|^{2/z_0})$
with $r_{0}$ the bare distance from the (clean) QCP.
To introduce quenched randomness, we dilute the system with
nonmagnetic impurities of spatial density $b$, i.e., we add a
potential, $\delta r(\mathbf{x})=\sum_{i}V[\mathbf{x}-\mathbf{x}(i)]$, to
$r_{0}$. Here, $\mathbf{x}(i)$ are the random positions of the impurities, and
$V(\mathbf{x})$ is a positive short-ranged impurity potential.

In our disordered LGW theory, quantum Griffiths phases occur in Ising systems ($n=1$) with
dissipationless dynamics ($z_0=1$) or in continuous-symmetry systems ($n>1$) with Ohmic
dissipation ($z_0=2$).\cite{Vojta06} We first focus on the Ising case,
minor differences for $n>1$ will be discussed later.

\begin{figure}
\includegraphics[width=7.5cm]{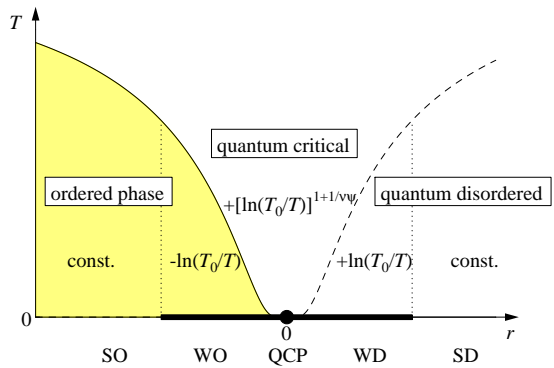}
\caption{(Color online:) Schematic phase diagram close to an infinite-randomness QCP.
   SO and SD denote the strongly ordered and disordered
   bulk phases while  WO and WD are the weakly ordered and disordered quantum Griffiths phases.
   The terms in the figure show the
   temperature-dependence of $\Gamma$ or $\Gamma_H$ (for pressure and field-tuned transitions,
   respectively).}
\label{Fig:pd_schematic}
\end{figure}

We start our analysis with the weakly disordered quantum Griffiths phase (WD in Fig.\
\ref{Fig:pd_schematic}).
Despite the dilution, there are large spatial regions devoid of impurities.
They can be locally in the magnetic phase even though the bulk system is still
nonmagnetic. The probability $w$ of finding such a rare region or cluster of linear size
$L_{RR}$ is exponentially small in its volume, $w \sim \exp(-b L_{RR}^d)$.
Because the cluster is locally ordered it acts as a two-level system with an energy gap
$\epsilon \sim \exp(-a L_{RR}^d)$. Combining the two exponential laws, we obtain
the well-known power-law density of states (see, e.g., Ref.\ \onlinecite{Vojta06}),
\begin{equation}
\rho(\epsilon) \sim \epsilon^{\lambda(r) - 1}~.
\label{eq:DOS_WD}
\end{equation}
We have parametrized the nonuniversal power law in terms of the Griffiths exponent $\lambda=b/a$.
It vanishes at the QCP and increases with
increasing distance $r$ from criticality. To determine the rare-region contribution to the
entropy at temperature $T$,
we note that each rare region with $\epsilon < T$ contributes an entropy of $\ln 2$
while those with $\epsilon > T$ do not contribute significantly.
Allowing for an $r$-dependent prefactor, we thus find
\begin{equation}
S(r,T) = N g(r)\, (T/T_0)^{\lambda(r)}~
\label{eq:S_WD}
\end{equation}
where $T_0$ is a microscopic temperature scale.

Thermal expansion coefficient and specific heat can now be calculated easily by taking
the appropriate derivatives of the entropy. From (\ref{eq:beta_def}) and
(\ref{eq:c_def}),
we find the leading low-temperature behavior to be
\begin{eqnarray}
\beta &=& \frac 1 {V_m p_c} g(r) \lambda'(r) \, (T/T_0)^{\lambda(r)} \ln(T_0/T)~,
\label{eq:beta_WD}\\
c_p &=& g(r) \lambda(r) \, (T/T_0)^{\lambda(r)}~,
\label{eq:c_WD}
\end{eqnarray}
where $\lambda'(r)$ denotes the derivative of $\lambda$ with respect to $r$.
In the Gr\"{u}neisen ratio, the temperature dependencies of $\beta$ and $c_p$
almost completely cancel, resulting in
\begin{equation}
\Gamma = \frac \beta {c_p} = \frac 1 {V_m p_c} \, \frac {\lambda'(r)}{\lambda(r)} \ln(T_0/T)~.
\label{eq:Gamma_WD}
\end{equation}
The rare region contribution to the Gr\"{u}neisen parameter diverges
logarithmically with decreasing temperature in the entire WD quantum Griffiths phase.
Because $\lambda$ increases with $r$, both the thermal expansion coefficient and
the Gr\"{u}neisen parameter are positive. This agrees with the notion that
the low-temperature entropy decreases with increasing distance from criticality.

In the weakly ordered (WO) quantum Griffiths phase, the relevant
degrees of freedom are strongly coupled clusters that are sufficiently isolated from the
(ordered) bulk system so that they can fluctuate independently. This requires that the
effective coupling $J_{\rm eff}$ of the cluster to the bulk is smaller than its energy gap
which still reads $\epsilon \sim \exp(-a L_{RR}^d)$. To isolate the cluster, it
must thus be surrounded by a large spatial region that is locally in the nonmagnetic phase.
Generically, the correlations will drop off exponentially with distance
in this region. The condition
$J_{\rm eff} < \epsilon$ thus implies that the \emph{linear} size of the isolating
region must vary as $\ln(1/\epsilon) \sim L_{RR}^d$ with the cluster size $L_{RR}$.
We conclude that the probability of finding a sufficiently isolated
cluster of size $L_{RR}$ drops off as $w \sim \exp[-\bar{b} (L_{RR}^{d})^d]$, i.e., much
faster than in the WD phase. The resulting density of states takes the form
\begin{equation}
\rho(\epsilon) \sim \frac 1 \epsilon
\exp\left[-\bar\lambda(r)\ln^d(\epsilon_0/\epsilon)\right]
\label{eq:DOS_WO}
\end{equation}
with a nonuniversal $\bar\lambda(r)$ which is the analog of the Griffiths exponent
$\lambda(r)$. Thus, we still obtain a gapless spectrum, but the singularity is weaker
than in the WD phase  in all
dimensions $d>1$. In particular, the density of states vanishes faster than any power law
with $\epsilon \to 0$. We
emphasize that (\ref{eq:DOS_WO}) is the generic result, special types of randomness
can lead to stronger singularities. For instance, in a percolation scenario (site
or bond dilution of a lattice), a shell of empty sites or bonds is sufficient to
completely isolate a
cluster. In this case, $w \sim \exp[-\bar{b} L_{RR}^{d-1}]$ and $\rho(\epsilon) \sim
\epsilon^{-1} \exp[-\bar\lambda(r)\ln^{1-1/d}(\epsilon_0/\epsilon)]$
giving rise to a singularity even stronger than in the WD phase.\cite{SenthilSachdev96}

The rare-region contribution to the entropy can be estimated as above by simply counting
the clusters with energy gap $\epsilon < T$. This gives
\begin{equation}
S(r,T) = N \bar g(r) \exp\left[ -\bar\lambda(r) \ln^d(T_0/T) \right ]~.
\label{eq:S_WO}
\end{equation}
Using (\ref{eq:beta_def}) and (\ref{eq:c_def}), we obtain the leading low-temperature
behavior of thermal expansion and specific heat,
\begin{eqnarray}
\beta &=& \frac { \bar g(r) \bar\lambda'(r)} {V_m p_c} \exp\left[ -\bar\lambda(r) \ln^d(T_0/T)
\right] \ln^d(T_0/T)~,~~
\label{eq:beta_WO}\\
c_p &=& d \bar g(r) \bar \lambda(r) \exp\left[ -\bar\lambda(r) \ln^d(T_0/T) \right ]
\ln^{d-1}(T_0/T)~.~~
\label{eq:c_WO}
\end{eqnarray}
This results in a Gr\"{u}neisen parameter of
\begin{equation}
\Gamma = \frac \beta {c_p} = \frac 1 {d V_m p_c} \, \frac {\bar\lambda'(r)}{\bar\lambda(r)}
\ln(T_0/T)~,
\label{eq:Gamma_WO}
\end{equation}
which is (except for the extra factor $1/d$) identical to the WD result (\ref{eq:Gamma_WD}).
Note that a different power of $\ln(\epsilon_0/\epsilon)$ in the exponent
of (\ref{eq:DOS_WO}) (as discussed above for percolation
disorder) would not change the temperature-dependence of $\Gamma$. In the WO quantum
Griffiths phase, $\bar\lambda$ decreases with increasing $r$ (approaching the QCP).
The thermal expansion coefficient and the Gr\"{u}neisen parameter are
therefore negative, again in agreement with the entropy decreasing with increasing
distance from criticality. Note that the rare region
contributions to both $\beta$ and $c_p$ in the WO phase are only weakly singular
for $d>1$. Therefore, they may be subleading
to contributions from other soft modes in the system,
making (\ref{eq:Gamma_WO}) hard to observe experimentally.


\emph{Scaling analysis}---We now complement the heuristic rare region theory
by a scaling analysis of the Gr\"{u}neisen parameter at infinite-randomness QCPs.
These exotic critical points emerge from the strong-disorder
renormalization group\cite{MaDasguptaHu79,Fisher92,IgloiMonthus05} and generally
occur in conjunction with the quantum Griffiths phases discussed above.\cite{Vojta06}

According to the strong-disorder renormalization group,
the density of independent clusters surviving at temperature $T$ scales like an inverse volume and
thus has the scaling form\cite{VojtaKotabageHoyos09}
\begin{equation}
n(r,T) = [\ln(T_0/T)]^{-d/\psi} \, \Phi\left ( r\, [\ln(T_0/T)]^{1/(\nu\psi)}\right)~.
\label{eq:n_scaling}
\end{equation}
where $\nu$ and $\psi$ are the correlation length and tunneling exponents.
The scaling function $\Phi(y)$ is analytic at $y\ge 0$ (because there is no
finite-temperature phase transition at $r\ge 0$). For small $y$, we can thus expand
$\Phi(y) = \Phi(0) + y \Phi'(0) + \ldots$. For large positive $y$ (in the WD quantum
Griffiths phase), $\Phi(y) = A y^{d\nu} \exp(-c x^{\nu\psi})$ with $A$ and $c$ constants.
The scaling function
$\Phi(y)$ has a singularity at some $y_c < 0$ marking the transition to the ordered phase.
This immediately gives the unusual form of the phase boundary, $T_c(r) = T_0
\exp[-(y_c/r)^{\nu\psi}]$, sketched in Fig.\ \ref{Fig:pd_schematic}.
Since the surviving clusters are essentially free, each contributes $s_0=\ln 2$
to the entropy. The scaling part of the entropy thus reads
\begin{eqnarray}
S(r,T) = N s_0 [\ln(T_0/T)]^{-d/\psi} \, \Phi\left (
r\,[\ln(T_0/T)]^{1/(\nu\psi)}\right)\,.~
\label{eq:S_scaling}
\end{eqnarray}

We first calculate the thermal expansion coefficient and the specific heat at criticality,
$r=0$. Applying (\ref{eq:beta_def}) and (\ref{eq:c_def}) to the scaling form (\ref{eq:S_scaling})
of the entropy, we obtain
\begin{eqnarray}
\beta &=& -\frac {s_0}{V_m p_c} \Phi'(0) [\ln(T_0/T)]^{-d/\psi+1/(\nu\psi)}~,
\label{eq:beta_QCP}\\
c_p &=& \frac{s_0 d}{\psi} \Phi(0) [\ln(T_0/T)]^{-d/\psi-1}~.
\label{eq:c_QCP}
\end{eqnarray}
Forming the ratio  $\beta/c_p$, we find that the critical part of
the Gr\"{u}neisen parameter behaves as
\begin{equation}
\Gamma = -\frac {\psi} {V_m p_c d} \, \frac {\Phi'(0)}{\Phi(0)}
\,[\ln(T_0/T)]^{1+1/(\nu\psi)}~.
\label{eq:Gamma_QCP}
\end{equation}
Equation (\ref{eq:Gamma_QCP}) holds in the entire quantum critical region $T > T_0
\exp[-|y_x/r|^{\nu\psi}]$ where the constant $y_x$ marks the crossover
of $\Phi(y)$.
The sign of $\Gamma$ does not follow from the scaling analysis, but because the
entropy accumulates close to the finite-temperature phase boundary, we generally expect
$\Gamma>0$ in the quantum critical region.\cite{GarstRosch05}

Let us now analyze the scaling form (\ref{eq:S_scaling}) of the entropy
in the WD quantum Griffiths phase, i.e., for $r>0$ and
$T < T_0 \exp[-(y_x/r)^{\nu\psi}]$. Using the large-argument limit
of the scaling function $\Phi(y)$,
the density of surviving clusters is given by $n(r,T) = A r^{d\nu}
\exp[-c r^{\nu\psi} \ln(T_0/T)]$. The resulting functional form of the entropy,
\begin{equation}
S(r,T) = N g(r)\, (T/T_0)^{\lambda(r)}~,
\label{eq:S_scaling_WD}
\end{equation}
is identical to that found in (\ref{eq:S_WD}) using
heuristic rare region arguments, but the scaling analysis also gives
$\lambda(r) = c r^{\nu\psi}$ and $g(r)= A s_0  r^{d\nu}$
in terms of the distance to criticality and the critical exponents. Inserting
$g(r)$ and $\lambda(r)$ into (\ref{eq:beta_WD}), (\ref{eq:c_WD}), and
(\ref{eq:Gamma_WD}) leads  to
\begin{eqnarray}
\beta &=& \frac {A s_0}{V_m p_c} r^{d \nu+\nu\psi -1} c\nu\psi (T/T_0)^{\lambda(r)}
\ln(T_0/T)~,
\label{eq:beta_WD_scaling}\\
c_p &=& A s_0 c \, r^{d\nu +\nu\psi} (T/T_0)^{\lambda(r)}~,
\label{eq:c_WD_scaling}\\
\Gamma &=& \frac 1 {V_m} \, \frac {\nu\psi}{p-p_c} \ln(T_0/T)~.
\label{eq:Gamma_WD_scaling}
\end{eqnarray}
The prefactor of the logarithmic temperature dependence of $\Gamma$
thus diverges
as $1/(p-p_c)$ at the QCP.

As discussed above, the behavior on the ordered side of the transition,
i.e., in the WO quantum Griffiths phase, is dimensionality and
disorder dependent. Once these are fixed,
the analysis can be performed in complete analogy to the WD quantum Griffiths
phase.


\emph{Conclusions}---We have determined the Gr\"{u}neisen parameter $\Gamma$
at pressure-tuned QPTs in the presence of quenched disorder. At an
infinite-randomness QCP, the critical contribution to
$\Gamma$ diverges as $[\ln(T_0/T)]^{1+1/(\nu\psi)}$ with $T\to 0$.
In the associated quantum Griffiths phases on both sides of the QCP, the
rare region contribution to $\Gamma$ behaves as $\ln(T_0/T)$  with a prefactor that diverges
and changes sign at criticality  ($\Gamma<0$ for $p<p_c$ and $\Gamma >0$ for $p>p_c$).
Our results must be contrasted with the behavior at clean QCPs, where $\Gamma$ diverges
as a power of $T$ at criticality but remains finite for all $p\ne p_c$.

In many systems, a QPT can be induced by doping instead of
pressure. If the main effect of doping is an expansion or compression of
the lattice, it acts as `chemical pressure'. Close to criticality, the effects of
pressure $p$ and doping $x$ can then be related via $(p-p_c)=c (x-x_c)$ with $c$ a constant.
Defining the distance from criticality for such a transition as $r=(x-x_c)/x_c$, this leads to
the relation $\partial/ \partial p = (c x_c)^{-1} \partial / \partial r$.
All our results for $\beta$ and $\Gamma$ thus hold if $c x_c$ is substituted for $p_c$.

If the transition is tuned by magnetic field instead of pressure, our calculations carry over
to the magnetocaloric effect $\Gamma_H$ defined in (\ref{eq:Gamma_H_def}). In fact, by replacing $V_m p_c$ by $H_c$
in the results for $\Gamma$, one obtains the corresponding expressions for $\Gamma_H$.
Note, however, that our analysis assumes random-$T_c$ type disorder which
does not break the order parameter symmetry. In magnetic-field
tuned transitions in the presence of disorder, stronger random-field type effects may
be generated.\cite{FishmanAharony79,TGKSF06,Schechter08,AnfusoRosch09} They would require a separate
analysis.

In our LGW theory, quantum Griffiths phases and infinite-randomness QCPs
occur either for Ising symmetry without dissipation or for continuous
$O(n)$ symmetry ($n>1$) and Ohmic dissipation.\cite{Vojta06} So far, we have focused
on the Ising case. The main difference for $n>1$ is that
ordered clusters act as (damped) quantum rotors rather than two-level systems.\cite{VojtaKotabageHoyos09}
In the Griffiths phase, this changes the prefactors
$g(r)$ and $\bar g(r)$ while the exponents $\lambda(r)$ and $\bar\lambda(r)$ remain
the same. At criticality,
the entropy (\ref{eq:S_scaling}) picks up an extra factor $\ln(T_0/T)$ from the entropy
of a single rotor. It drops out in the ratio $\beta/c_p$. Thus, our results remain valid
in the $O(n)$ case, at least at criticality and in the WD Griffiths phase. In the
WO Griffiths phase, the rare region contributions to $\beta$ and $c_p$ will be
overcome by conventional soft mode terms (because the dissipative $O(n)$ system is gapless).

We now turn to experiment. CePd$_{1-x}$Rh$_{x}$ features a ferromagnetic QPT that
appears to be dominated by rare regions.\cite{SWKCGG07,Westerkampetal09}
The phase boundary develops a tail characteristic of a smeared QPT,
and at temperatures above the tail, magnetization, susceptibility and specific heat
display nonuniversal power-laws as expected in a quantum Griffiths phase.
Recent measurements of the thermal expansion\cite{Westerkampetal09} resulted in a
very weakly temperature-dependent Gr\"{u}neisen parameter close to the
putative transition at $x_c \approx 0.87$, in qualitative agreement with our theory.
However, the variation of $\Gamma$ with doping $x$ differs considerably
from our results. This may be caused by the fact that the doping
is not isoelectronic. It thus not only acts as chemical pressure by inducing a lattice
compression (as assumed in our discussion), but it also changes the electronic structure
directly. To disentangle these effects one could prepare a sample with
doping close to $x_c$ and then drive it through the QPT by
pressure.


We acknowledge discussion with M.\ Brando, M. Deppe, P.\ Gegenwart, and T. Westerkamp.
This work has been supported in part by the NSF under grant
no. DMR-0339147, by Research Corporation, and by the University of Missouri Research
Board.

\bibliographystyle{apsrev}
\bibliography{../00Bibtex/rareregions}
\end{document}